\documentclass{svmult}
\usepackage{makeidx}
\usepackage{graphicx}
\usepackage{multicol}

\makeindex

\begin{document}

\title*{The Structure of Light Nuclei and Its Effect on Precise Atomic 
Measurements}
\titlerunning{The Structure of Light Nuclei}

\author{James L.\ Friar}
\institute{Theoretical Division, Los Alamos National Laboratory, Los Alamos, NM,
USA \texttt{friar@lanl.gov}}

\label{04}
\maketitle

\vspace*{-2.5in}
\hfill \fbox{\parbox[t]{0.95in}{LA-UR-02-6889}}
\vspace*{2.2in}

\begin{abstract}
This review consists of three parts: (a) what every atomic physicist needs to
know about the physics of light nuclei; (b) what nuclear physicists can do for
atomic physics; (c) what atomic physicists can do for nuclear physics. A brief
qualitative overview of the nuclear force and calculational techniques for light
nuclei will be presented, with an emphasis on debunking myths and on recent
progress in the field. Nuclear quantities that affect precise atomic
measurements will be discussed, together with their current theoretical and
experimental status. The final topic will be a discussion of those atomic
measurements that would be useful to nuclear physics, and nuclear calculations
that would improve our understanding of existing atomic data.
\end{abstract}

\section{Introduction}

\begin{quote} 
\em ....numerical precision is the very soul of science.....
\end{quote}

This quote\cite{darcy} from Sir D'Arcy Wentworth Thompson, considered by many to
be the first biomathematician, could well serve as the motto of the field of
precise atomic measurements, since precision is the {\it raison d'\^etre} of
this discipline. I have always been in awe of the number of digits of accuracy
achievable by atomic physics in the analysis of simple atomic
systems\cite{2S1S}.  Nuclear physics, which is my primary field and interest,
must usually struggle to achieve three digits of numerical significance, a level
that atomic physics would consider a poor initial effort, much less a decent
final result.

The reason for the differing levels of accuracy is well known: the theory of
atoms is QED, which allows one to calculate properties of few-electron systems
to many significant figures\cite{review}. On the other hand, no aspect of
nuclear physics is known to that precision.  For example, a significant part of
the ``fundamental'' nuclear force between two nucleons must be determined
phenomenologically by utilizing experimental information from nucleon-nucleon
scattering\cite{2gen}, very little of which is known to better than 1\%. In
contrast to that level of precision, energy-level spacings in few-electron atoms
can be measured so precisely that nuclear properties influence significant
digits in those energies\cite{d-p}.  Thus these experiments can be interpreted
as either a measurement of those nuclear properties, or corrections must be
applied to eliminate the nuclear effects so that the resulting measurement tests
or measures non-nuclear properties. That is the purview of this review.

The single most difficult aspect of a calculation for any theorist is assigning
uncertainties to the results.  This is not always necessary, but in calculating
nuclear corrections to atomic properties it is essential to make an effort. That
is just another way to answer the question,``What confidence do we have in our
results?'' Because it is important for atomic physicists to be able to judge
nuclear results to some degree, this discussion has been slanted towards answers
to two questions that should be asked by every atomic physicist. The first is:
``What confidence should I have in the values of nuclear quantities that are
required to analyze precise atomic experiments?'' The second question is: ``What
confidence should I have that the nuclear output of my atomic experiment will be
put to good use by nuclear physicists?''

\section{Myths of Nuclear Physics}

Every field has a collection of myths, most of them being at least partially
true at one time.  Myths propagate in time and distort the reality of the
present.  A number of these are collected below, some of which the author once
believed. The resolution of these ``beliefs'' also serves as a counterpoint to
the very substantial progress made in light-nuclear physics in the past 15
years, which continues unabated.

My myth collection includes:

$\bullet$  The strong interactions (and consequently the nuclear force) aren't
well understood, and nuclear calculations are therefore unreliable.

$\bullet$ Large strong-interaction coupling constants mean that perturbation
theory doesn't converge, implying that there are no controlled expansions in
nuclear physics.

$\bullet$  The nuclear force has no fundamental basis, implying that
calculations are not trustworthy.

$\bullet$  You cannot solve the Schr\"odinger equation accurately because of the
complexity of the nuclear force.

$\bullet$  Nuclear physics requires a relativistic treatment, rendering a
difficult problem nearly intractable.

All of these myths had some (even considerable) truth in the past, but today
they are significant distortions of our current level of knowledge.

\section{The Nuclear Force}

Most of the recent progress in understanding the nuclear force is based on a
symmetry of QCD\index{QCD}, which is believed to be the underlying theory of the
strong interactions (or an excellent approximation to it).  It is generally the
case that our understanding of any branch of physics is based on a framework of
symmetry principles. QCD has ``natural'' degrees of freedom (quarks and gluons)
in terms of which the theory has a simple representation.  The (strong) chiral
symmetry\index{Chiral~symmetry!in~QCD} of QCD results when the quark masses
vanish,  and is a more complicated analogue of the chiral symmetry that results
in QED\index{Chiral~symmetry!in~QED} when the electron mass vanishes.  The
latter symmetry explains, for example, why (massless or high-energy) electron
scattering from a spherical (i.e., spinless) nucleus vanishes in the backward
direction.

The problem with this attractive picture is that it does not involve the degrees
of freedom most relevant to experiments in nuclear physics: nucleons and pions.
It is nevertheless possible to ``map'' QCD (expressed in terms of quarks and
gluons) into an ``equivalent'' or surrogate theory expressed in terms of nucleon
and pion degrees of freedom.  This surrogate works effectively only at low
energy.  The small-quark-mass symmetry limit becomes a small-pion-mass symmetry
limit.  In general this (slightly) broken-symmetry theory has $m_{\pi} c^2 \ll
\Lambda$, where the pion mass is $m_{\pi} c^2 \cong$ 140 MeV and $\Lambda \sim$
1 GeV is the mass scale of QCD bound states (heavy mesons, nucleon resonances,
etc.). The seminal work on this surrogate theory, now called chiral perturbation
theory\index{Chiral~perturbation~theory} (or $\chi$PT), was performed by Steve
Weinberg\cite{QCD}, and many applications to nuclear physics were pioneered by
his student, Bira van Kolck\cite{nQCD}. From my perspective they demonstrated
two things that made an immediate impact on my understanding of nuclear
physics\cite{pc}: (1) There is an alternative to perturbation theory in coupling
constants, called ``power counting\index{Power~counting},'' that converges
geometrically like $(Q/\Lambda)^N$, where $Q \sim m_{\pi}c^2$ is a relevant
nuclear energy scale, and the exponent $N$ is constrained to have $N \geq 0$;
(2) nuclear physics mechanisms are severely constrained by the chiral symmetry.
These results provide nuclear physics with a well-founded rationale for
calculation.

\begin{figure}   \centering   
\includegraphics[scale=0.85,bb= 100 347 478 600]{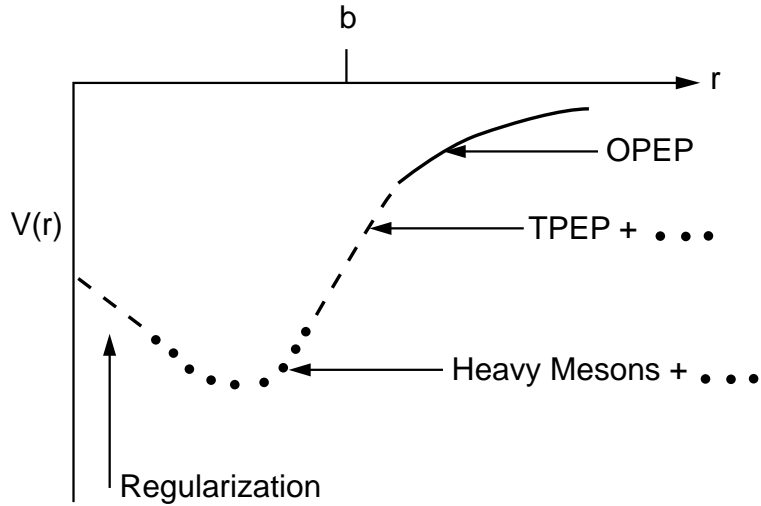}   
\caption{Cartoon of the nuclear potential, $V (r)$, showing regions of  
importance}
\end{figure}

This scheme divides the nuclear-force regime in a natural way into a long-range
part\index{Nuclear~force!long-range} (which implies a low energy, $Q$, for the
nucleons) and a short-range\index{Nuclear~force!short-range} part (corresponding
to high energy, $Q$, between nucleons). This is indicated in Fig.~(1), which is
a cartoon of the potential between two nucleons meant only to indicate
significant regions and mechanisms. Since $\chi$PT is effective only at low 
energies, we expect that only the long-range part of the nuclear force can be 
treated successfully by utilizing only the pion degrees of freedom. This would 
be the region with $r > b$. We need to resort to
phenomenology\index{Nuclear~phenomenology} (i.e., fitting to nucleon-nucleon
scattering data) to treat systematically the short-range part of the interaction
($r < b$).

The long-range nuclear force is calculated in much the same way that atomic
physics calculates the interactions in an atom using QED.  Both are illustrated
in Fig.~(2). The dominant interaction between two nucleons is the exchange of a
single pion illustrated in Fig.~(2b) (One-Pion-Exchange Potential or
``OPEP''\index{Nuclear~force!OPEP}) and denoted $V_{\pi}$. Its atomic analogue
is one-photon exchange in Fig.~(2a) (containing the dominant Coulomb force).
Because it is such an important part of the nuclear potential, it is fair to
call $V_{\pi}$ the ``Coulomb force'' of nuclear physics.  Smaller contributions
arise from the two-pion-exchange potential in Fig.~(2e) (called
``TPEP''\index{Nuclear~force!TPEP}), which is the analogue of two-photon
exchange between charged particles shown in Fig.~(2d). There is even an analogue
of the atomic polarization force in Fig.~(2g), where two electrons 
simultaneously polarize their nucleus using their electric fields. The nuclear
analogue involving three nucleons simultaneously is illustrated in Figs.~(2h)
and (2i), and is called a three-nucleon
force\cite{3NF}\index{Nuclear~force!three-nucleon}. Although relatively
weak compared to $V_{\pi}$ (a few percent), three-nucleon forces play an
important role in fine-tuning nuclear energy levels. The final ingredient is an
important short-range interaction\index{Nuclear~force!short-range} (which
must be determined by phenomenology) shown in Fig.~(2c) that has no direct
analogue in the physics of light atoms. Just as one can exchange three photons,
three-pion-exchange is possible and is depicted in Fig.~(2f).

It is worth recalling that the uncertainty principle tells us that exchanging
light particles produces longer-range forces and exchanging heavier particles
produces shorter-range forces. Thus OPEP has a longer range than TPEP, as
illustrated in Fig.~(1). Many mechanisms have been proposed for the short-range
part of the nuclear force, such as heavy-meson exchange, for example. Any
meson-exchange mechanism produces singular forces, which are regularized to make
them finite. However one chooses to do this, the part of the nuclear force
inside $b$ must be adjusted to fit the nucleon-nucleon scattering
data\index{Nuclear~phenomenology}, and no individual parameterization of the
short-range force is intrinsically superior (i.e., it doesn't matter how you do
it).

\begin{figure} \centering 
\includegraphics[scale=0.70]{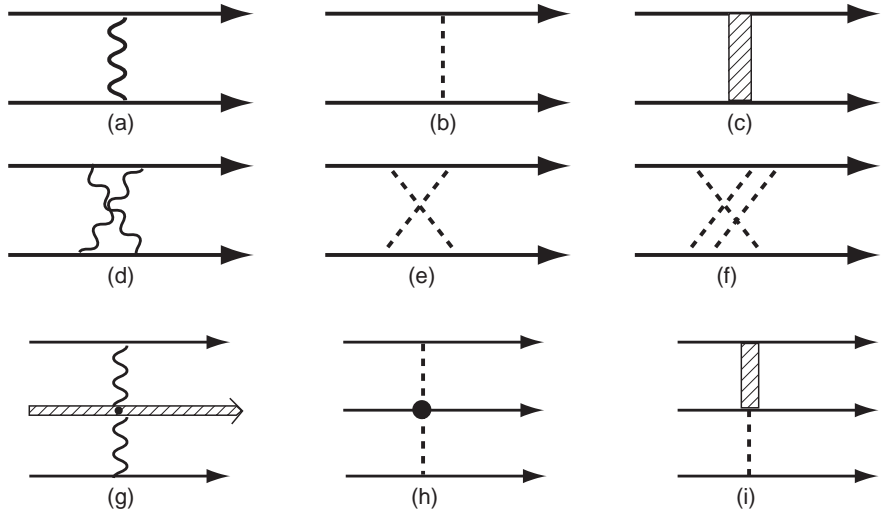}  
\caption{First- and second-order (in $\alpha$, the fine structure constant)  
atomic interactions resulting from photon exchange are shown in the left-most  
column, where solid lines are electrons, wiggly lines are photons, and the  
shaded line is a nucleus. The analogous nuclear interactions resulting from  
pion exchange are shown in the middle column, where solid lines are nucleons  
and dashed lines are pions. Nuclear processes involving short-range interactions
(shaded vertical areas) are shown in the right-most column, together with a
three-pion-exchange interaction}
\end{figure}

\begin{figure} \centering 
\includegraphics[totalheight=3.5in]{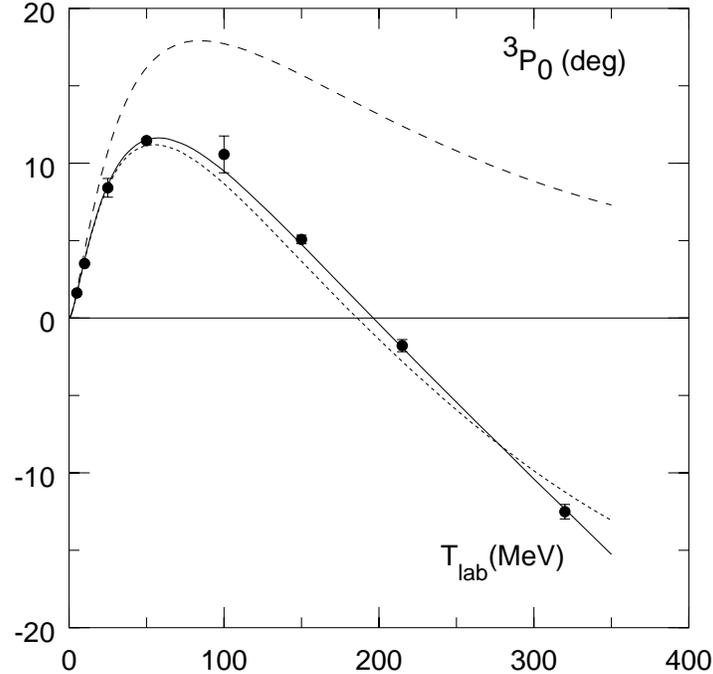}  
\caption{$^3P_0$ phase shift (in degrees) calculated only with the OPEP tail for
$r > b$ (dashed line), and with one (dotted) and three (solid)
short-range-interaction terms added. The experimental results are indicated by
separate points with error bars\protect\cite{psa}}
\end{figure}

How all of this works in practice is indicated in Fig.~(3). Imagine that you
throw away {\it all} of the nuclear potential inside $r = b$ (with $b$ chosen to
be 1.4 fm) in Fig.~(1), keeping only the tail of the force between two nucleons.
Now compute a phase shift (the $^3$P$_0$, for example). This very modest physics
input predicts the basic shape of the phase shift (dashed line) as a function of
energy. This variation with energy is a consequence of the small pion mass
(compared to the energy scale in the figure). What is missing in this curve is a
smooth (negative) short-range contribution that grows roughly in proportion to
the energy. We can fill in the missing short-range interaction inside $r = b$ by
adding a potential term specified by one short-range parameter. This produces
the dotted line, which is a rather good fit, and adding two more terms (solid
line) produces a nearly perfect fit to the experimental results. Fixing the
short-range\index{Nuclear~force!short-range} part of the potential looks very
much like making an effective-range expansion. All useful physics is specified
by a few parameters, and the details are completely unimportant.

What are the consequences of exchanging a pion rather than a photon?  The
pseudoscalar nature of the pion mandates its spin-dependent coupling to a 
nucleon, and this leads to a dominant tensor force\index{Nuclear~force!tensor}
between two nucleons.  Except for its radial dependence, the form of $V_{\pi}$
mimics the interaction between two magnetic dipoles, as seen in the Breit
interaction, for example. Thus we have in nuclear physics a situation that is
the converse of the atomic case:  a dominant tensor force and a smaller central
force.  In order to grasp the difficulties that nuclear physicists face, imagine
that you are an atomic physicist in a universe where magnetic (not electric)
forces are dominant, and where QED can be solved only for long-range forces and
you must resort to phenomenology to generate the short-range part of the force
between electrons and nuclei.

Although this may sound hopeless, it is merely difficult. The key to handling
complexities is adequate computing power, and that became routinely available
only in the late 1980s or early 1990s. Since then there has been explosive
development in our understanding of light nuclei.  Underlying all of these
developments is an improved understanding of the nuclear force.  It is
convenient to divide nuclear forces and their history into three distinct time
periods.

{\it First-generation}\index{Nuclear~force!first-generation} nuclear forces were
developed prior to 1993. They all contained the one-pion-exchange force, but
everything else was relatively crude. The fits to the nucleon-nucleon scattering
data (needed to parameterize the short-range part of that force) were
indifferent.

{\it Second-generation}\index{Nuclear~force!second-generation} forces were
developed beginning in 1993\cite{2gen}. They were more sophisticated and
generally very well fit to the scattering data.  As an example of how well the
fitting worked, the Nijmegen group (which pioneered this sophisticated
procedure) allowed the pion mass to vary in the Yukawa function defining
$V_{\pi}$\index{Nuclear~force!OPEP}, and then fit that mass. They also allowed
different masses for the neutral and charged pions that were being exchanged and
found\cite{pi-mass}

\begin{equation} m_{\pi^\pm} = 139.4(10)\, {\rm MeV} \, , \end{equation}
\begin{equation} m_{\pi^{0}}\, = 135.6(13)\, {\rm MeV} \, , \end{equation} both
results agreeing with free pion masses ($m_{\pi^{\pm}} = 139.57018(35)$ MeV and
$m_{\pi^0} = 134.9766(6)$ MeV \cite{PDG}). It is both heartening and a bit
amazing that the masses of the pions can be determined to better than 1\% using
data taken in reactions that have no free pions!  This result is the best
quantitative proof of the importance of pion degrees of freedom in nuclear
physics.

{\it Third-generation}\index{Nuclear~force!third-generation}  nuclear forces are
currently under development. These forces are quite sophisticated and
incorporate two-pion exchange\index{Nuclear~force!TPEP}, as well as $V_{\pi}$. 
All of the pion-exchange forces (including three-nucleon
forces)\index{Nuclear~force!three-nucleon} are being generated in accordance 
with the rules of chiral perturbation theory\index{Chiral~perturbation~theory}.
One expects even better fits to the scattering data. This is clearly work in
progress, but preliminary calculations and versions have already
appeared\cite{3gen}.

\section{Calculations of Light Nuclei}

\begin{figure}[p]
  \includegraphics[scale=0.63]{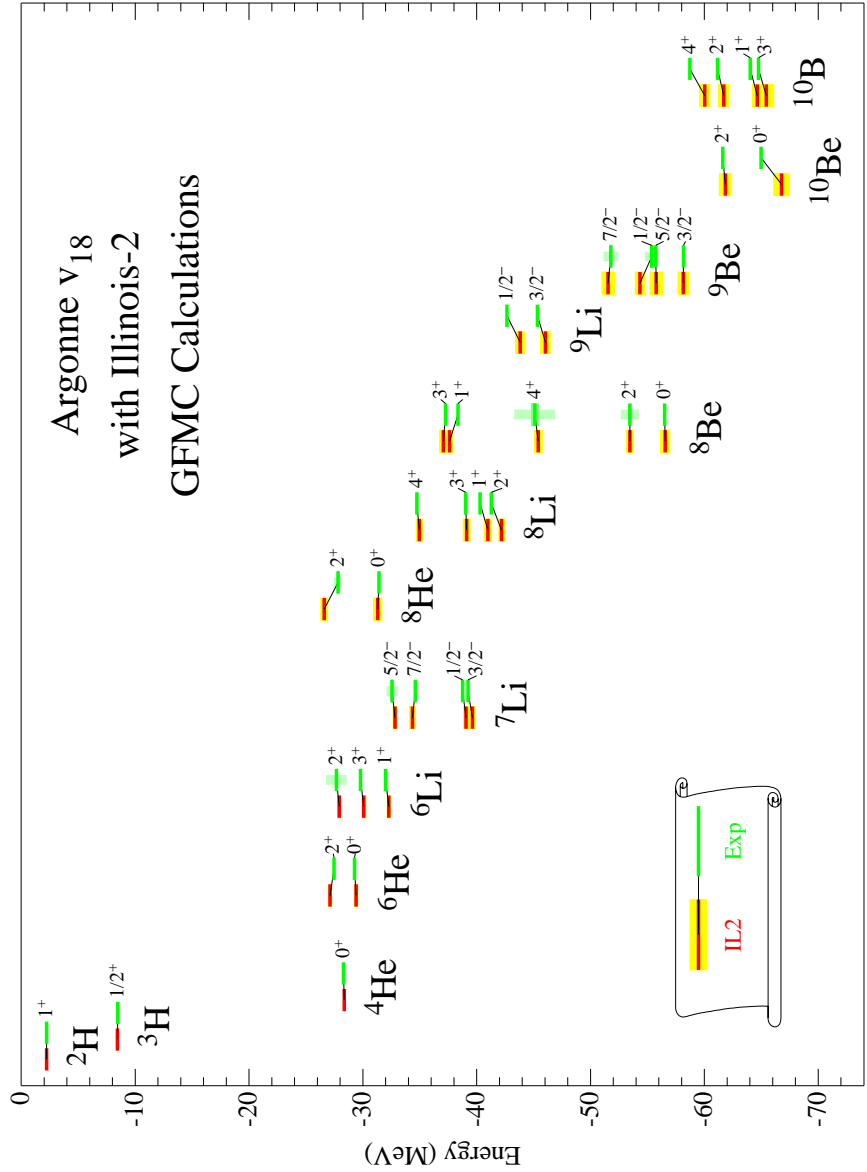}
  \caption{GFMC calculations of the binding energies of the levels (labelled by 
   their spins and parities) of light nuclei with as many as ten nucleons. These
   calculations use a common Hamiltonian and have a numerical uncertainty on
   the order of 1\%. Heavy shaded lines to the left are calculated energies 
   (with errors), while light shaded lines to the right are experimental 
   energies. The label ``IL2'' refers to the Illinois-2 model of the 
   three-nucleon force that is used in all of the calculations together with 
   the Argonne V$_{18}$ two-nucleon force}
\end{figure}

Having a nuclear force is not very useful unless one can calculate nuclear
properties with it.  Such calculations are quite difficult. Until the middle
1980s only the two-nucleon problem had been solved with numerical errors smaller
than 1\%.  At that time the three-nucleon systems $^3$H and $^3$He were
accurately
calculated\index{Accurate~nuclear~calculations!trinucleon@$^3$H~and~$^3$He}
using a variety of first-generation nuclear-force models\cite{3N}.  Soon
thereafter the
$\alpha$-particle\index{Accurate~nuclear~calculations!alpha@$^4$He} ($^4$He) was
calculated by Joe Carlson, who pioneered a technique that has revolutionized our
understanding of light nuclei: Green's Function Monte Carlo
(GFMC)\cite{GFMC}\index{Green's~Function~Monte~Carlo}.

The difficulty in solving the Schr\"odinger equation for nuclei is easily
understood, although it was not initially obvious. Nuclei are best described in
terms of nucleon degrees of freedom. Nucleons come in two types, protons and
neutrons, which have nearly the same masses and can be considered as the up and
down components of an ``isospin'' degree of freedom. If one also includes its
spin, a single nucleon thus has four internal degrees of freedom. Two nucleons
consequently have 16 internal degrees of freedom, which is roughly the number of
components in the nucleon-nucleon force (coupling spin, isospin and orbital
motion in a very complicated way).  To handle this complexity one again requires
fast computers, and that is a fairly recent development.

The GFMC\index{Green's~Function~Monte~Carlo} technique has been used to solve 
for all of the bound (and some unbound) states of nuclei with up to 10 nucleons.
One member of this collaboration (Steve Pieper\cite{steve}) calculated that the
ten-nucleon Schr\"odinger equation requires the solution of more than 200,000
coupled second-order partial-differential equations in 27 continuous variables,
and this can be accomplished with numerical errors on the order of 1\%! A subset
of the results of this impressive calculation are shown in Fig.~(4)\cite{10A}.

Although the nucleon-nucleon scattering data alone can predict the binding
energy of the deuteron ($^2$H) to within about 1/2\%, the experimental binding
energy is used as input data in fitting the nucleon-nucleon potential. The
nuclei $^3$H and $^3$He (not shown) are slightly underbound without a
three-nucleon force\index{Nuclear~force!three-nucleon}, and that force can be
adjusted to remedy the underbinding. This highlights both the dominant nature of
the nucleon-nucleon force and the relative smallness of three-nucleon forces,
which is nevertheless appropriate in size to account for the small discrepancies
that result from using only nucleon-nucleon forces in calculations of nuclei
with more than two nucleons.

Once the $^3$H binding energy is fixed, the binding energy of $^4$He is then
accurately predicted to within about 1\%. The five-nucleon systems (not shown)
are unbound, but their properties are rather well reproduced. The six-nucleon
systems are also well predicted.  There are small problems with more
neutron-rich nuclei (compare $^9$Li with $^7$Li or $^8$He with $^6$He or
$^4$He), but only 3 adjustable parameters in the three-nucleon force allow
several dozen energy levels to be quite well 
reproduced\cite{10A}\index{Accurate~nuclear~calculations!a10@$A=$ 2-10}. Because
nuclei are weakly bound systems, there are large cancellations between the
(large) potential and (large) kinetic energies, leaving small binding energies.
The results shown in Fig.~(4) are quite remarkable, especially given that small
(fractional) discrepancies in the energy components lead to large effects on the
binding energies.

We note finally that power counting\index{Power~counting} can be used to show
that light nuclei are basically non-relativistic, and relativistic corrections
are on the order of a few percent.  Power counting is a powerful qualitative
technique for determining the relative importance of various mechanisms in
nuclear physics.

\section{What Nuclear Physics Can Do for Atomic Physics}

With our recently implemented computational skills we in nuclear physics can
calculate many properties of light nuclei with fairly good accuracy.  This is
especially true for the deuteron, which is almost unbound and is computationally
simple.  Although nuclear experiments don't have the intrinsic accuracy of
atomic experiments, many nuclear quantities that are relevant to precise atomic
experiments can also be measured using nuclear techniques, and usually with
fairly good accuracy.

What quantities are we talking about?  The nuclear length
scale\index{Nuclear~scales} is set by $R \sim 1$ fm $= 10^{-5}$ \AA.  The much
larger atomic length scale of $a_0 \sim $ 1 \AA\ means that an expansion in
powers of $R/a_0$ makes great sense, and a typical wavelength for an atomic
electron is so large compared to the nuclear size that only moments of the
nuclear observables come into play. This also corresponds to an expansion in
$\alpha$, the fine-structure constant, and $m_{\rm e} R$, where $m_{\rm e}$ is
the electron mass. This is a rapidly converging series.

For processes that have nuclear states inside loops (such as polarizabilities)
the excitation energies of those states play a significant role. Although states
of any energy can be excited in principle, in practice the effective energy of
(virtual) excitation for light nuclei (call it $\bar{\omega}_N$) is within a
factor of two of 10~MeV, except for the more tightly bound $\alpha$-particle,
which also has a smaller radius as a consequence. This number follows from
the uncertainty principle and the fact that nucleons in a light nucleus
have a radius of about 2~fm. The deuteron's weak binding generates the lowest
values, which is about 6 MeV for the deuteron's
electric\index{Electric~polarizability!deuterium}
polarizability\index{Polarizability!electric} ($\alpha_{\rm E} \sim
1/\bar{\omega}_N$). Using the value of $\bar{\omega}_N$ = 10 MeV, we find
$\bar{\omega}_N R \sim 1/10$, which is a reasonably small expansion parameter.

\begin{table}[ht]
\centering
\caption{Orders in $\alpha$ where various contributions to the Lamb shift for
S-states have been calculated. The label ``f.s.'' denotes a contribution from
nuclear finite size or nuclear structure. Once nuclear physics enters a
process at a given order, higher orders will also have nuclear corrections. A
``$\mbox{$-$}$'' indicates that although a  complete calculation of nuclear
contributions has not been made, such contributions are expected. Names refer
to the person who first calculated the leading-order term of that type.
References and the meanings of other labels are given in the text}
\begin{tabular}{l l l l l}
\hline \noalign{\smallskip}
Process \hspace{0.5in} & $\alpha^2$ & $\alpha^4$ & $\alpha^5$ & $\alpha^6$ \\ 
\noalign{\smallskip}
\hline
NR Coulomb & Bohr \hspace{0.2in}& f.s.    & f.s.      & f.s.     \\ \hline
Rel. Coulomb    & & Dirac     &           & f.s.     \\ \hline
Recoil          & & Darwin \hspace{0.02in}&\mbox{$-$} &\mbox{$-$}\\ \hline
Nucl. Structure & &           & f.s.      &\mbox{$-$}\rule{0in}{2.5ex}\\ \hline 
Vacuum Pol.     & &           & Uehling \hspace{0.2in}& f.s.     \\ \hline
Radiative       & &           & Bethe     & f.s.     \\ \hline
\end{tabular}
\end{table}

At what levels do various nuclear mechanisms affect the Lamb
shift\index{Lamb~shift!nuclear~finite~size}? The (lowest) orders in $\alpha$
that receive nuclear contributions (for S-states) are sketched in Table~(1).
The various mechanisms are divided into static Coulomb (both non-relativistic
and relativistic), recoil (inverse powers of the nuclear mass, $M$), nuclear
structure, vacuum polarization, and radiative processes. The nuclear effects are
conveniently divided into two categories: those that directly involve only the
properties of the nuclear ground state, and those that involve virtual excited
states and are traditionally called ``nuclear structure.'' A radius is a good
example of the former, while a polarizability is the prototype of the latter.
Calculational techniques are quite different for these two categories.

It is beyond the scope of this review to list detailed formulae and extensive
references to past work. I strongly recommend the recent review of
\cite{review}, which is extremely well organized. An entire section is devoted
to nuclear contributions, and these are listed in their Table~(10) with
references and numerical values for the hydrogen atom. A sketch of how these
contributions scale is given below together with some of the more recent
references.

The leading-order non-relativistic energy is simply the Bohr energy of order
$\alpha^2$. Nuclear finite-size\index{Lamb~shift!nuclear~finite~size}
contributions of non-relativistic type (i.e., generated by the Schr\"odinger
equation) begin for S-states in order $(Z \alpha)^4$\cite{KKS} and are
proportional to $R^2$; they have also been calculated in order $(Z \alpha)^5$
and $(Z \alpha)^6$\cite{fs1,fs2}. The Dirac energy has a leading-order $(Z
\alpha)^4$ term, while the nuclear finite-size contributions of relativistic
type begin in order $(Z \alpha)^6$ and are proportional to $R^2$. A recent
calculation exists for deuterium\cite{ho-size}. The non-relativistic finite-size
corrections of order $(Z \alpha)^5$ and $(Z \alpha)^6$ are tiny for electronic
atoms (they contain higher powers of $m_{\rm e} R$, which is very small), but
are not necessarily small for muonic atoms ($m_{\mu} R$ is about 1 for most 
light nuclei), which was the original motivation for developing them. P-state
finite-size effects begin in order $(Z \alpha)^6$ and are of both relativistic
($\sim R^2$) and non-relativistic ($\sim R^4$) types.

The most important nuclear-structure mechanism is the electric
polarizability\index{Electric~polarizability} (which has a long history and will
be discussed in more detail later), and this generates a leading contribution of
order $\alpha^2 (Z \alpha)^3$. Coulomb corrections
\index{Electric~polarizability!Coulomb~corrections} of order $\alpha^2 (Z
\alpha)^4$ were developed in the context of a greatly simplified model of the
polarizability in muonic atoms\cite{He4-x} (which would not be applicable to
electronic atoms).

The Uehling mechanism for vacuum polarization is of order $\alpha (Z \alpha)^4$,
while the first nuclear corrections are of order $\alpha (Z
\alpha)^5$\cite{fs-vp,fs-vp-h,fs-rad, fs-rad-p}. The leading-order radiative
process is also of order $\alpha (Z \alpha)^4$, while the nuclear finite-size
corrections begin in order $\alpha (Z \alpha)^5$\cite{fs-rad,fs-rad-p}. Both of
these nuclear corrections are proportional to $R^2$. Recoil corrections have a
long and interesting history that predates the Schr\"odinger equation (C.\ G.\
Darwin derived the leading term of order $(Z \alpha)^4$ using Bohr-Sommerfeld
quantization; see the references in \cite{breit}). To the best of my knowledge
no published calculation exists for the nuclear-finite-size recoil corrections,
which begin in order $(Z \alpha)^5$, although the techniques of \cite{GY} lead 
to a result proportional to $R^2/M$, which should be very small. We note finally
the hadronic vacuum polarization\index{Vacuum~polarization!hadronic}, which
(although not nuclear in origin) is generated by the strong
interactions\cite{had-vp}.

One quantity through which nuclear size manifests itself is the nuclear charge
form factor (the Fourier transform of the nuclear ground-state charge
density\index{Nuclear~charge~density}, $\varrho$), which is given by
\pagebreak

\begin{equation}
F (\vec{q}) = \int d^3 r \, \varrho (\vec{r}) \, \exp (i \vec{q} \cdot \vec{r}) 
\cong Z (1- \frac{\vec{q}^2}{6}\langle r^2 \rangle_{\rm ch} + \cdots \, ) 
- \textstyle{1\over2} \, \vec{q}^{\alpha} \vec{q}^{\beta}\, Q^{\alpha \beta} + 
\cdots \, ,
\end{equation}
where $\vec{q}$ is the momentum transferred from an electron to the nucleus,
$Q^{\alpha \beta}$ is the nuclear quadrupole-moment tensor, $Z$ is the total
nuclear charge, and $\langle r^2 \rangle_{\rm ch}$ is the mean-square radius of
the nuclear charge density. These moments should dominate the nuclear 
corrections to atomic energy levels because $|\vec{q}|$ in an atom is set by 
the (very small) atomic scales. Using $F$ to construct the electron-nucleus 
Coulomb interaction, one obtains

\begin{equation}
V_{\rm C} ( \vec{r} ) \cong - \frac{Z \alpha}{r} + \frac{2 \pi Z \alpha}{3} 
\langle r^2 \rangle_{\rm ch} \, \delta^3 (\vec{r}) - \frac{Q \alpha}{2 r^3} 
\frac{( 3 \, (\vec{S} \cdot \hat{\vec{r}})^2 -\vec{S}^2)}{S (2 S-1)} + 
\cdots \, ,
\end{equation}
where $\vec{S}$ is the nuclear spin operator and $Q$ is the nuclear quadrupole
moment\index{Nuclear~quadrupole~moment!Coulomb~interaction} (which vanishes
unless the nucleus has spin $S \geq 1$). The Fourier transform of the nuclear
ground-state current density\index{Nuclear~current~density}  has a similar
expansion

\begin{equation}
\vec{J} ( \vec{q} ) = \int d^3 r \  \vec{J} (\vec{r}) \, \exp ( i \vec{q} 
\cdot \vec{r} ) \cong -i \vec{q} \times \mbox{\boldmath $\mu$} \, (1- 
\frac{\vec{q}^2}{6} \langle r^2 \rangle_{\rm M}\, + \, \cdots \, ) \; + \, 
\cdots \, ,
\end{equation}
where $\mbox{\boldmath{$\mu$}}$ is the nuclear magnetic
moment\index{Magnetic~moment!nuclear} and $\langle r^2 \rangle_{\rm M}$ is the
mean-square radius\index{Nuclear~radii} of the magnetization
density\index{Nuclear~radii!magnetic}. The first term generates the usual atomic
hyperfine interaction.

\begin{table}
\centering
\caption{Values of the root-mean-square charge and magnetic radii and the
quadrupole moment (if nonvanishing) of the nucleons and various light nuclei
obtained by nuclear experiments, together with a selected reference. If two
values are given, the second value is that obtained by an atomic or molecular
measurement}
\begin{tabular}{l l l l l l l}
\hline \noalign{\smallskip}
Nucleus \hspace{0.1in} 
&$\langle r^2 \rangle_{\rm ch}^{1/2}\, ({\rm fm})$\hspace{0.1in} 
&ref.\hspace{0.3in} 
& $\langle r^2 \rangle_{\rm M}^{1/2}\, ({\rm fm})$ \hspace{0.2in}
&ref. \hspace{0.15in} 
&Q\, (fm$^2$) \hspace{0.3in} 
& ref.  \\ \noalign{\smallskip} \hline
{ }H   & 0.880\,(15)&\cite{Coulomb}&0.836\,(9)&\cite{fit}  &\mbox{$-$}  & 
\rule{0in}{2.5ex} \\ 
      & 0.883\,(14)&\cite{2loop} &           &            &\mbox{$-$}  & \\ 
\hline
$^2$H & 2.130\,(10)&\cite{ST}    & 2.072\,(18)&\cite{ingo}&0.282\,(19) &
\cite{Qnuc} \rule{0in}{2.5ex} \\ 
      &            &             &           &            &0.2860\,(15)&
\cite{Qhd1,Qhd2} \rule{0in}{2.5ex} \\ \hline
$^3$H & 1.755\,(87)&\cite{ingo}  & 1.84\,(18)&\cite{ingo} &\mbox{$-$}  & 
\rule{0in}{2.5ex} \\ \hline
$^3$He& 1.959\,(34)&\cite{ingo}  & 1.97\,(15)&\cite{ingo} &\mbox{$-$}  &
\rule{0in}{2.5ex} \\
      & 1.954\,(8) &\cite{shiner}&           &            &\mbox{$-$}  & \\ 
\hline
$^4$He& 1.676\,(8) &\cite{sickx} & \mbox{$-$}&            &\mbox{$-$}  &
\rule{0in}{2.5ex}\\ \hline \hline
\noalign{\smallskip}
Nucleon & $\langle r^2 \rangle_{\rm ch}^{ }\, ({\rm fm}^2)$ &ref. & 
$\langle r^2 \rangle_{\rm M}^{1/2}\, ({\rm fm})$ &ref. & \rule{0in}{2.5ex} \\
\noalign{\smallskip} \hline
  n   &\mbox{$-$}0.1140\,(26)&\cite{neutron}&0.873\,(11)&\cite{nmag}& &
\\ \hline
\end{tabular}
\end{table}

Electron-nucleus scattering experiments\index{Electron-nucleus~scattering} are
the primary technique used to measure moments of nuclear charge and current 
densities that are relevant to atomic physics\cite{ingo}, and some appropriate
values of these quantities are tabulated in Table~(2). An exception is the 
measurement of the deuteron's quadrupole 
moment\index{Nuclear~quadrupole~moment!deuterium} ($Q = 0.282(19)$ fm$^2$) 
obtained by scattering polarized deuterons from a high-Z nuclear target 
at low energy\cite{Qnuc}.  This result is consistent with the molecular
determination ($Q = 0.2860(15)$ fm$^2$)\cite{Qhd1,Qhd2}, but its error is an 
order of magnitude larger. Although there is no reason to believe that the 
(tensor) electric polarizability\index{Electric~polarizability!tensor} of the
deuteron\cite{tau} plays a significant role in the H-D (molecular)
quadrupole-hyperfine splitting\index{Hyperfine~splitting!quadrupole} that was
used to determine $Q$, that correction was not included in the analysis. It was
included in the analysis of the nuclear measurement.

I highly recommend the recent review of electron-nucleus
scattering\index{Electron-nucleus~scattering} by Ingo Sick\cite{ingo}, which
contains values of the charge and magnetic radii of light nuclei.  That review
not only lists the best and most recent values of quantities of interest, but
discusses reliability and technical details for those who are interested. One
result from that review is listed in Table~(2) and is important for the
discussion below.  The errors of the tritium ($^3$H)
radii\index{Nuclear~radii!3H@$^3$H} are nearly an order of magnitude larger than
those of deuterium.  Of all the light nuclei tritium is the most poorly known
experimentally, although the charge radius can now be calculated with reasonable
accuracy.

\begin{figure}
  \centering
  \includegraphics[scale=0.9]{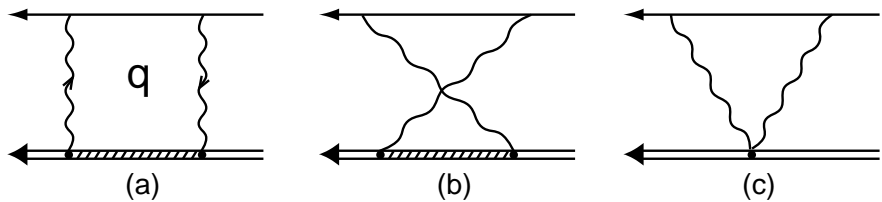}
  \caption{The direct two-photon process is shown in (a), the crossed-photon
   process in (b), and ``seagull'' contributions in (c). The seagulls reflect 
   non-nucleonic processes and terms necessary for gauge invariance. In these 
   graphs the double lines represent a nucleus, the single lines an electron, 
   the wiggly lines a (virtual) photon. The shading represents the set of 
   nuclear excited states. The loop momentum is $q$, and integrating over this 
   momentum sets the scales of the nuclear part of the process}
\end{figure}

In addition to moments of the nuclear charge and current densities, various
components and moments of the nuclear Compton
amplitude\index{Compton~amplitude!nuclear} can play a significant role.
Mechanisms that contribute to the polarizabilities are shown in Fig.~(5). The
direct (sequential) exchange of photons and the crossed-photon process are shown
in (a) and (b), while the ``seagull'' process is shown in (c). The latter
mechanism is required by gauge invariance in any model of hadrons with
structure. The exchange of pions between nucleons generates such terms, for
example\cite{c-mec}\index{Compton~amplitude!meson-exchange~currents}. Because
these are loop diagrams, they involve an integral over all momenta ($q$), and
this sets the nuclear scales of the problem. The nuclear size 
scale\index{Nuclear~scales} ($R$), the electron mass ($m_{\rm e}$), and the
average virtual-excitation energy ($\bar{\omega}_N$, appropriate to the shaded
part of the line in (a) and (b) that indicates excited nuclear states) determine
the generalized polarizabilities\cite{ho-pol}\index{Polarizability!generalized}.
The process is dominated by the usual electric\index{Polarizability!electric} 
and magnetic\index{Polarizability!magnetic} polarizabilities and their
logarithmic
modifications\cite{d-pol}\index{Polarizability!logarithmic~modification}.

Specific examples are the (scalar) electric polarizability, $\alpha_{\rm E}$,
and the nuclear spin-dependent
polarizability\index{Polarizability!spin-dependent} ($\sim \vec{S}$). The latter
interacts with the electron spin to produce a contribution to the
electron-nucleus hyperfine
splitting\index{Hyperfine~splitting!spin-dependent~polarizability}.  There
exists a recent calculation of the latter for
deuterium\cite{d-nu}\index{Hyperfine~splitting!deuterium}. Values of the nuclear
electric polarizability for light
nuclei\index{Electric~polarizability!light~nuclei} obtained from calculations or
experiments are listed in Table~(3). There are either calculations or
measurements of $\alpha_{\rm E}$ for $^2$H\cite{d-pol,d-pol-g,d-pol-x},
$^3$H\cite{3pol} and $^3$He\cite{3pol,4pol,he3-pol,rinker}, and
$^4$He\cite{He4-x,He4-t}\index{Lamb~shift!MU@muonic $^4$He}. With the exception
of the $^3$He experimental results, there is reasonable consistency. The
decreasing size of the polarizabilities for heavier nuclei is caused by their
increased binding. The $\alpha$-particle has more than 10 times the binding
energy of the deuteron, and its polarizability is an order of magnitude smaller.

\begin{table}[ht]
\centering
\caption{Values of the electric polarizability of light nuclei, both theoretical
and experimental, where the latter have been determined by nuclear experiments.
No uncertainties were given for the $^3$H, $^3$He, and $^4$He calculations in
\cite{3pol,4pol}, but they are likely to be smaller than about 10\%. The 
$^4$He result was used in analyses of muonic He\protect\cite{He4-x,He4-t}}
\begin{tabular}{l l l l l}
\hline \noalign{\smallskip}
Nucleus \hspace{0.2in}
& $\alpha_{\rm E}^{\rm calc}\, ({\rm fm}^3)$ \hspace{0.2in}
&ref. \hspace{0.2in}
&$\alpha_{\rm E}^{\rm exp}\, ({\rm fm}^3)$ \hspace{0.2in}
&ref. \\ \noalign{\smallskip} \hline
$^2$H  &0.6328\,(17)&\cite{d-pol}  & 0.61\,(4)   &\cite{d-pol-g} \\ 
       &            &              & 0.70\,(5)   &\cite{d-pol-x} \\ \hline
$^3$H  & 0.139      &\cite{3pol}   &\mbox{$-$}   & \rule{0in}{2.5ex} \\ \hline
$^3$He & 0.145      &\cite{4pol}   & 0.250\,(40) &\cite{he3-pol}\\
       &            &              & 0.130\,(13) &\cite{rinker} \\ 
\hline
$^4$He & 0.076      &\cite{4pol}   & 0.072\,(4)  &\cite{He4-x}
\rule{0in}{2.5ex}\\ \hline 
\end{tabular}
\end{table}

The physics of hyperfine splittings is in general rather different from the
physics that contributes to the Lamb shift. It should therefore be no surprise
that the nuclear physics that contributes to hyperfine splittings is also
quite different; it is also more complicated than its Lamb-shift counterpart.
The dominant nuclear physics that we discussed previously was the physics of the
nuclear charge density\index{Nuclear~charge~density}, in the form of moments of
the static charge density (i.e., radii) and electric dipole moments that
contribute to polarizabilities.

The primary nuclear mechanism in hyperfine splittings is the magnetic
interaction caused by the nuclear magnetization
density\index{Nuclear~current~density!magnetization density}. This density
is not as well understood as the charge density. The primary reason is that
the same mesons whose exchange binds nuclei together also contribute to the
nuclear currents if they carry a charge. The pions that we discussed earlier
generate a very important component of that current\cite{c-mec}. The reason for
the dichotomy between nuclear charges and currents can be understood by 
imagining that charged-meson exchange between nucleons is instantaneous. In this
limit we know that the transmitted charge is always on a nucleon. In other words
only nucleon degrees of freedom matter, which is the normal situation for the 
charge density. The power counting\index{Power~counting} that we discussed 
earlier states that corrections to the nuclear charge operator are small 
($\sim$ 1\%), and include a type that vanishes for instantaneous meson 
exchanges. That is not the case for the current, however, since any flow of
charge (even from a virtual meson) produces a current that is not simply related
to nucleon degrees of freedom, and that current can couple to photons. These 
meson-exchange currents\index{Nuclear~current~density!meson-exchange~currents} 
(often denoted ``MEC'') can be as large as those generated by the usual nuclear
convection current. Various tricks can be used to eliminate part of our 
ignorance, but the nuclear current density is less well understood than the 
nuclear charge density. Atomic hyperfine splittings provide us with an excellent
opportunity to learn about nuclear currents in a very different setting.

Although most of the hyperfine experiments in light
atoms\index{Hyperfine~splitting!light~atoms} were performed decades ago, there
has recently been renewed theoretical interest, and the accuracy of the QED
calculations is sufficient to extract nuclear information\cite{sav-nu}. The 
differences between the QED calculations and the experimental results can be 
interpreted as nuclear corrections, and those are significant, as indicated in
Table~(4). The S-state results in this table (presented as a ratio) have been 
taken from Table~(1) of the recent work of Ivanov and Karshenboim\cite{sav-nu}.

\begin{table}[htb]
\centering
\caption{Difference between hyperfine experiments and QED hyperfine calculations
for the $n\underline{\rm th}$ S-state of light hydrogenic atoms times $n^3$,
expressed as parts per million of the Fermi energy. This difference is
interpreted as nuclear contributions to the hyperfine
splitting\protect\cite{sav-nu}. A negative entry indicates that the theoretical
prediction without nuclear corrections is too large}
\begin{tabular}{l l l l l}
\hline \noalign{\smallskip}
\multicolumn{5}{c}{$n^3 (E_{\rm hfs}^{\rm exp} - 
E_{\rm hfs}^{\rm QED})/E_{\rm F}\, {\rm (ppm)}$} \\ \noalign{\smallskip} \hline
 State \hspace{0.3in} & H \hspace{0.3in}    & $^2$H \hspace{0.3in} 
 & $^3$H \hspace{0.3in}    & $^3$He$^+$ \rule{0in}{2.5ex}\\ \hline \hline
 1S    & -33   & 138    & -38       & 222        \\ \hline
 2S    & -33   & 137    &\mbox{$-$} & 221        \\ \hline 
\end{tabular}
\end{table}

Hyperfine structure is generated by short-range interactions. The dominant Fermi
contribution ($E_{\rm F}$) arises from a $\delta$-function, and that produces a
dependence on the square of the electron's $n\underline{\rm th}$ S-state wave
function at the origin, $|\phi_n (0)|^2$, which is proportional to 1/n$^3$. Most
nuclear effects have the same dependence ($\sim 1/n^3$), which has been removed
from the results in Table~(4). The 1S and 2S results are seen to be consistent
at this level of accuracy, with 1S experimental results typically being much
more accurate.

More calculations of the nuclear contributions to hyperfine splittings in light
atoms\index{Hyperfine~splitting!light~atoms} are badly needed if we are to use
this information to learn about the currents in light nuclei. These
contributions come in the form of Zemach
moments\cite{zemach}\index{Hyperfine~splitting!Zemach~moments} (ground-state
quantities) and spin-dependent
polarizabilities\index{Hyperfine~splitting!spin-dependent~polarizability}
(discussed above). There exists a considerable literature on the latter subject
dating back 50 years. The recent work of Mil'shtein and Khriplovich\cite{d-nu}
has pointed out a serious defect in that older work. Although the leading-order
terms are essentially non-relativistic in origin (for the nucleons in a 
nucleus), the sub-leading-order terms are not, and require relativity (for the 
nucleons) in order to obtain a correct result. This is not terribly surprising,
since the same physics that enters that polarizability also enters the
Gerasimov-Drell-Hearn sum rule\cite{GDH}\index{Gerasimov-Drell-Hearn~sum~rule},
which requires relativity at the nucleon level\cite{GDH-nuc}, and is a topic of
considerable current interest in nuclear physics\cite{henry}. The calculations
of \cite{d-nu} suggest that deuterium\index{Hyperfine~splitting!deuterium} at
least can be understood using fairly simple nuclear models. This needs to be
checked using more sophisticated models. We note that the Zemach
correction\cite{zemach}\index{Hyperfine~splitting!Zemach~moments} adds to the
ratio in Table~(4), improving the agreement between experiment and theory for H
and $^3$H. The large positive value of that ratio for deuterium suggests a large
polarizability correction, which is confirmed by \cite{d-nu}.

\section{The Proton Size}

One recurring problem in the hydrogen Lamb shift is the appropriate value of the
mean-square radius\index{Nuclear~radii!proton} of the proton, $\langle r^2
\rangle_{\rm p}$, to use in calculations. Some older determinations\cite{HMW} 
disagree strongly with more recent ones\cite{Simon}. As shown in (3), the slope
of the charge form factor (with respect to $\vec{q}^2$) at $\vec{q}^2$ = 0
determines that quantity. The form factor is measured by scattering electrons
from the proton at various energies and scattering
angles\index{Electron-nucleus~scattering}.

There are (at least) four problems associated with analyzing the
charge-form-factor data to obtain the proton size.  The first is that the
counting rates in such an experiment are proportional to the flux of electrons
times the number of protons in the target seen by each electron.  That product
must be measured.  In other words the measured form factor for low $\vec{q}^2$ 
is ($a - b \frac{\vec{q}^2}{6} + \cdots$), where $b/a = \langle r^2 \rangle_{\rm
p}$. The measured normalization $a$ (not exactly equal to 1) clearly influences
the value and error of $\langle r^2 \rangle_{\rm p}$. Most analyses
unfortunately don't take the normalization fully into account, and \cite{norm}
estimates that a proper treatment of the normalization of available data could
increase $\langle r^2 \rangle_{\rm p}^{1/2}$ by about 0.015 fm and increase its
error, as well. In an atom, of course, the normalization is precisely 
computable.

Another source of error is neglecting higher-order corrections in $\alpha$
(i.e., Coulomb
corrections)\index{Electron-nucleus~scattering!Coulomb~corrections}. and
\cite{Coulomb} demonstrates that this increases $\langle r^2 \rangle_{\rm
p}^{1/2}$ by about 0.010 fm. A similar problem in analyzing deuterium data was
resolved in \cite{ST}. Another difficulty that existed in the past was a lack of
high-quality low-$\vec{q}^2$ data.  The final problem is that one must use a
sufficiently flexible fitting function to represent $F ( \vec{q})$, or the
errors in the radius will be unrealistically low. All of the older analyses had
one or more of these flaws.

Most of the recent analyses\cite{Simon,Coulomb,fit} are compatible if the
appropriate corrections are made. An analysis by Rosenfelder\cite{Coulomb}
contains all of the appropriate ingredients, and he obtains $\langle r^2
\rangle_{\rm p}^{1/2}$ = 0.880(15) fm. There is a PSI experiment now underway to
measure the Lamb shift in muonic hydrogen\index{Lamb~shift!muonic~hydrogen},
which would produce the definitive result for $\langle r^2 \rangle_{\rm p}$
\cite{PSI,savely}. One expects the results of that experiment to be compatible
with Rosenfelder's result. Extraction of the proton radius\cite{2loop} from the
electronic Lamb shift\index{Lamb~shift!nuclear~finite~size} is now somewhat
uncertain because of controversy involving the two-loop diagrams. These diagrams
are significantly less important in muonic hydrogen, where the relative roles of
the vacuum polarization and radiative processes are reversed.

\section{What Atomic Physics Can Do for Nuclear Physics}

The single most valuable gift by atomic physics to the nuclear physics community
would be the accurate determination of the proton mean-square radius: $\langle
r^2 \rangle_{\rm p}$.  This quantity is important to nuclear theorists who wish
to compare their nuclear wave function calculations with measured mean-square
radii.  In order for an external source of electric field (such as a passing
electron) to probe a nucleus, it is first necessary to ``grab'' the proton's
intrinsic charge distribution\index{Nuclear~radii!proton}, which then maps out
the mean-square radius of the proton probability
distribution\index{Nuclear~radii!wave~function} in the wave function: $\langle
r^2 \rangle_{\rm wfn}$. Thus the measured mean-square radius of a nucleus,
$\langle r^2 \rangle$, has the following components:

\begin{equation}
\langle r^2 \rangle = \langle r^2 \rangle_{\rm wfn} + \langle r^2 \rangle_{\rm 
p}+ \frac{N}{Z} \langle r^2 \rangle_{\rm n} + \frac{1}{Z} \langle r^2 
\rangle_{\ldots} \, ,
\end{equation}
where the intrinsic contribution of the N neutrons has been included as well as
that of the Z protons, and $\langle r^2 \rangle_{\ldots}$ is the contribution of
everything else, including the very interesting (to nuclear physicists)
contributions from strong-interaction mechanisms and relativity in the nuclear
charge density\cite{czech}\index{Nuclear~charge~density!exotic~contributions}.
Because the neutron looks very much like a positively charged core surrounded by
a negatively charged cloud, its mean-square radius\index{Nuclear~radii!neutron}
has the opposite sign to that of the proton, whose core is surrounded by a
positively charged cloud. It should be clear from (6) that $\langle r^2
\rangle_{\rm p}$ (which is much larger than $\langle r^2 \rangle_{\rm n}$) is an
important part of the overall mean-square radius. Its present uncertainty
degrades our ability to test the wave functions of light nuclei.

\begin{figure}
  \centering
  \includegraphics[scale=1.0]{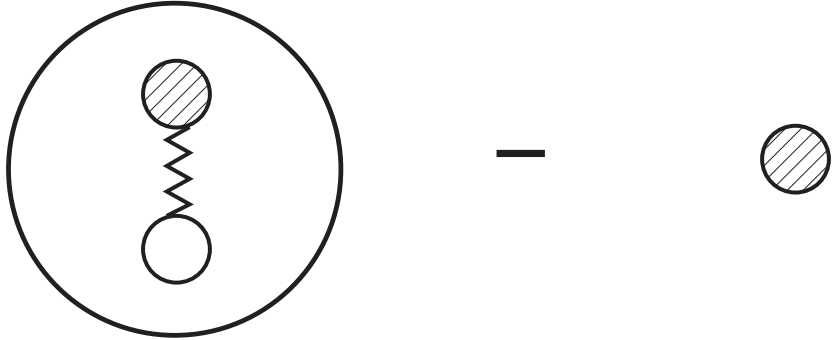}
  \caption{Cartoon of the $^2$H - H isotope shift, illustrating how the effect 
  of the finite size of the proton (shaded small circle) in deuterium is 
  cancelled in the measurement. The finite size of the neutron (open small 
  circle) and the electromagnetic interaction mediated by the strong-interaction
  (binding) mechanism (indicated by the jagged line between the nucleons) do 
  affect the deuteron's charge radius (see text)}
\end{figure}

The next most important measurements are isotope
shifts\index{Isotope~shift!light~atoms} in light atoms or ions. Since isotope
shifts measure differences in frequencies for fixed nuclear charge Z, the effect
of the protons' intrinsic size cancels in the difference. This is particularly
important given the current lack of a precise value for the proton's radius. The
neutrons' effect is relatively small and can be rather easily eliminated, and
thus one is directly comparing differences in wave functions, or of small
contributions from $\langle r^2 \rangle_{\ldots}$. Isotope shifts are therefore
especially ``theorist-friendly'' measurements, since they are closest to
measuring what nuclear theorists actually calculate.

Precise isotope-shift measurements have been performed for $^4$He -
$^3$He\cite{shiner}\index{Isotope~shift!helium~isotopes} and for $^2$H - $^1$H
(D-H)\cite{d-p}\index{Isotope~shift!hydrogen~isotopes}. A measurement of $^6$He
- $^4$He is being undertaken\cite{ANL} at ANL.  Gordon Drake has written about
and strongly advocated such measurements in the Li
isotopes\cite{Li-IS}\index{Isotope~shift!lithium~isotopes}. These are all highly
desirable measurements. Because the $^3$H (tritium) charge radius currently has
large errors, in my opinion the single most valuable measurement to
be undertaken for nuclear physics purposes would be the tritium-hydrogen ($^3$H
- $^1$H) isotope shift.  An extensive series of calculations using
first-generation nuclear forces found $\langle r^2 \rangle_{\rm wfn}^{1/2}$ for
tritium to be 1.582(8) fm\index{Nuclear~radii!3H@$^3$H}, where the
``error'' is a subjective estimate\cite{radius}. This number could likely be
improved by using second-generation nuclear forces, although it will never be as
accurate as the corresponding deuteron value, which we discuss next.

The D-H isotope shift\index{Isotope~shift!deuterium-hydrogen} in the 2S-1S
transition reported by the Garching group\cite{d-p} was

\begin{equation}
\varDelta \nu = 670 \ \, 994 \ \, 334.64 (15) \ {\rm kHz} \, .
\end{equation}
\hspace*{1.71in} $\uparrow$ \hspace{0.25in} $\uparrow$ \hspace{0.05in} 
$\uparrow$ \hspace{0.01in} $\uparrow$ \\ Most of this effect is due to the
different masses of the two isotopes (and begins in the first significant
figure, indicated by an arrow). The precision is nevertheless sufficiently high
that the mean-square-radius\index{Nuclear~radii!deuterium} effect in the sixth
significant figure (second arrow) is much larger than the error.  The electric
polarizability of the deuteron\index{Electric~polarizability!deuterium}
influences the eighth significant figure, while the deuteron's magnetic
susceptibility\index{Magnetic~susceptibility!deuterium} contributes to the tenth
significant figure.  It becomes difficult to trust the interpretation of the
nuclear physics at about the 1 kHz level, so improving this measurement probably
wouldn't lead to an improved understanding of the nuclear physics.

Analyzing this isotope shift and interpreting the residue (after applying all
QED corrections) in terms of the deuteron's radius leads to the
results\cite{iso} in Table~(5). The very small binding energy of the deuteron
produces a long wave function tail outside the nuclear potential (interpretable
as  a proton cloud around the nuclear center of mass), which in turn leads to an
easy and very accurate calculation of the mean-square radius of the (square of
the) wave function\index{Nuclear~radii!wave~function}. Subtracting this
theoretical radius from the experimental deuteron radius (corrected for the
neutron's size\index{Nuclear~radii!neutron}) determines the effect of $\langle
r^2 \rangle_{\ldots}$ on the radius. Although this difference is quite small, it
is nevertheless significant and half the size of the error in the corresponding
electron-scattering measurement (see Table~(2)). The high-precision analysis in
Table~(5) of the content of the deuteron's charge radius would have been
impossible without the precision of the atomic D-H isotope-shift measurement.
This measurement has given nuclear physics unique insight into small mechanisms
that are at present poorly understood\cite{MEC}.

\begin{table}[htb]
\centering
\caption{Theoretical and experimental deuteron radii for pointlike nucleons.
The deuteron wave function radius corresponding to second-generation nuclear
potentials and the experimental point-nucleon charge radius of the deuteron
(i.e., with the neutron charge radius removed) are shown in the first two
columns, followed by the difference of experimental and theoretical results. The
difference of the experimental radius with and without the neutron's size is
given last for comparison purposes\protect\cite{neutron}}
\begin{tabular}{l l l l}
\hline \noalign{\smallskip}
$          \langle r^2 \rangle_{\rm wfn}^{1/2}\, ({\rm fm})$ \hspace{0.1in}
& \rule{0in}{2.5ex} $_{\rm exp}\langle r^2 \rangle_{\rm pt}^{1/2}\, ({\rm fm})$ 
\hspace{0.2in}
&${\rm difference}\, ({\rm fm})$ \hspace{0.2in}
&$\varDelta \langle r^2 \rangle_{\rm n}^{1/2}\, ({\rm fm})$ \\ 
\noalign{\smallskip} 
\hline
1.9687(18) \rule{0in}{2.5ex}& 1.9753(10) & 0.0066(21) & \mbox{$-$}0.0291(7)\\ 
\hline
\end{tabular}
\end{table}

\section{Summary and Conclusions}

Nuclear forces and nuclear calculations in light nuclei are under control in a
way never before attained.  This progress has been possible because of the great
increase in computing power in recent years. Many of the nuclear quantities that
contribute to atomic measurements have been calculated or measured to a
reasonable level of accuracy, a level that is improving with time.  Isotope
shifts are valuable contributions to nuclear physics knowledge, and are
especially useful to theorists who are interested in testing the quality of
their wave functions for light nuclei. In special cases such as deuterium these
measurements provide the only insight into the size of small contributions to
the electromagnetic interaction that are generated by the underlying
strong-interaction mechanisms. In my opinion the tritium-hydrogen isotope
shift\index{Isotope~shift!tritium-hydrogen} would be the most useful measurement
of that type. One especially hopes that the ongoing PSI experiment is successful
in measuring the proton size via the Lamb shift in muonic hydrogen.

\section{Acknowledgements}
The work of J.\ L.\ Friar was performed under the auspices of the United States
Department of Energy. The author would like to thank Savely Karshenboim for his
interest in nuclear aspects of precise atomic measurements and for giving me the
opportunity to discuss this problem.

\label{04_}
\begin{theindex}

  \item Accurate~nuclear~calculations
    \subitem $A=$ 2-10, 9
    \subitem $^4$He, 7
    \subitem $^3$H~and~$^3$He, 7

  \indexspace

  \item Chiral~perturbation~theory, 3, 7
  \item Chiral~symmetry
    \subitem in~QCD, 3
    \subitem in~QED, 3
  \item Compton~amplitude
    \subitem meson-exchange~currents, 13
    \subitem nuclear, 13

  \indexspace

  \item Electric~polarizability, 11
    \subitem Coulomb~corrections, 11
    \subitem deuterium, 10, 18
    \subitem light~nuclei, 14
    \subitem tensor, 13
  \item Electron-nucleus~scattering, 13, 16
    \subitem Coulomb~corrections, 16

  \indexspace

  \item Gerasimov-Drell-Hearn~sum~rule, 16
  \item Green's~Function~Monte~Carlo, 7, 9

  \indexspace

  \item Hyperfine~splitting
    \subitem deuterium, 14, 16
    \subitem light~atoms, 15, 16
    \subitem quadrupole, 13
    \subitem spin-dependent~polarizability, 14, 16
    \subitem Zemach~moments, 16

  \indexspace

  \item Isotope~shift
    \subitem deuterium-hydrogen, 18
    \subitem helium~isotopes, 18
    \subitem hydrogen~isotopes, 18
    \subitem light~atoms, 17
    \subitem lithium~isotopes, 18
    \subitem tritium-hydrogen, 19

  \indexspace

  \item Lamb~shift
    \subitem muonic $^4$He, 14
    \subitem muonic~hydrogen, 17
    \subitem nuclear~finite~size, 10, 11, 17

  \indexspace

  \item Magnetic~moment
    \subitem nuclear, 12
  \item Magnetic~susceptibility
    \subitem deuterium, 18

  \indexspace

  \item Nuclear~charge~density, 11, 14
    \subitem exotic~contributions, 17
  \item Nuclear~current~density, 12
    \subitem magnetization density, 14
    \subitem meson-exchange~currents, 15
  \item Nuclear~force
    \subitem first-generation, 7
    \subitem long-range, 3
    \subitem OPEP, 4, 7
    \subitem second-generation, 7
    \subitem short-range, 3--5
    \subitem tensor, 6
    \subitem third-generation, 7
    \subitem three-nucleon, 4, 7, 9
    \subitem TPEP, 4, 7
  \item Nuclear~phenomenology, 3, 5
  \item Nuclear~quadrupole~moment
    \subitem Coulomb~interaction, 12
    \subitem deuterium, 13
  \item Nuclear~radii, 12
    \subitem $^3$H, 13, 18
    \subitem $^3$He, 22
    \subitem deuterium, 18
    \subitem magnetic, 12
    \subitem neutron, 17, 19
    \subitem proton, 16, 17
    \subitem wave~function, 17, 19
  \item Nuclear~scales, 10, 14

  \indexspace

  \item Polarizability
    \subitem electric, 10, 14
    \subitem generalized, 14
    \subitem logarithmic~modification, 14
    \subitem magnetic, 14
    \subitem spin-dependent, 14
  \item Power~counting, 3, 9, 15

  \indexspace

  \item QCD, 3

  \indexspace

  \item Vacuum~polarization
    \subitem hadronic, 11

\end{theindex}
\end{document}